\documentclass[12pt,a4paper]{article}
\usepackage{graphicx}
\usepackage{wrapfig}
\begin{document}
\begin{center}
{\bfseries  Light Scalar Mesons as Manifestation of Spontaneously
Broken Chiral Symmetry\,\footnote{{\small Talk at The
International Bogolyubov Conference "Problems of Theoretical and
Mathematical Physics" devoted to the 100th anniversary of
N.N.Bogolyubov's birth  that was held from August 21 to August 22,
2009 in Moscow at the Russian Academy of Sciences (RAS) and from
August 23 to August 27, 2009 in Dubna at the Joint Institute for
Nuclear Research (JINR). }}} \vskip 5mm N.N. Achasov \vskip 5mm
{\small {\it Sobolev Institute for Mathematics, 630090
Novosibirsk, Russia }\footnote{{\small E-mail:
achasov@math.nsc.ru}}}
\end{center}

\vskip 5mm \centerline{\bf Abstract} \vskip 1mm
 Attention is paid to the
production mechanisms of light scalars that reveal their nature.\\
 We  reveal the chiral shielding of the $\sigma(600)$
meson. We show that the kaon loop mechanism of the $\phi$
radiative decays, ratified by experiment, points to the four-quark
nature of light scalars. We show also that  the light scalars are
produced in the two photon collisions via four-quark transitions
in contrast to the classic $P$ wave tensor $q\bar q$ mesons that
are produced via two-quark transitions $\gamma\gamma\to q\bar q$.
The   history of spontaneous breaking of symmetry in quantum
physics is discussed in Appendix.

 \vskip 2mm

\noindent\textbf{1\ \ Introduction}

 The scalar channels
in the region up to 1 GeV became a stumbling block of QCD. The
point is that both perturbation theory and sum rules do not work
in these channels because there are not solitary resonances in
this region.

As  Experiment suggests, in chiral limit confinement forms
colourless observable hadronic fields and spontaneous breaking
 of chiral symmetry with massless pseudoscalar fields.
There are two possible scenarios for QCD realization at low
energy:\\ 1. $U_L(3)\times U_R(3)$ linear $\sigma$ model,\\ 2.
$U_L(3)\times U_R(3)$ non-linear $\sigma$ model.\\ The
experimental nonet of the light scalar mesons suggests
$U_L(3)\times U_R(3)$ linear $\sigma$ model.

\vskip 2mm

\noindent\textbf{1\ \ \boldmath{$SU_L(2)\times SU_R(2)$} Linear
$\sigma$ Model \cite{GL60}, Chiral Shielding  in $\pi\pi\to\pi\pi$
\cite{AS9407}}

Hunting the light  $\sigma$ and $\kappa$ mesons had begun in the
sixties. But the fact that both $\pi\pi$ and $\pi K$ scattering
phase shifts do not pass over $90^0$ at putative resonance masses
prevented to prove their existence in a conclusive way.

Situation changes when we showed that in the $SU_L(2)\times
SU_R(2)$ linear $\sigma$ model \cite{GL60} there is a negative
background phase which hides the $\sigma$ meson \cite{AS9407}. It
has been made clear that shielding wide lightest scalar mesons in
chiral dynamics is very natural. This idea was picked up and
triggered new wave of theoretical and experimental searches for
the $\sigma$ and $\kappa$ mesons.

Our approximation is as follows (see Fig.\,1):\\[5pt]
$T^0_0$\,=\,$\frac{T_0^{0(tree)}}{1-i\rho_{\pi\pi}
T_0^{0(tree)}}$\,=\,$\frac{e^{2i(\delta_{bg}+\delta_{res})}-1}{2i\rho_{\pi\pi}}
$\,=\,$\frac{e^{2i\delta^0_0}-1}{2i\rho_{\pi\pi}}$ = $T_{bg}+
e^{2i\delta_{bg}}T_{res}$,\\[5pt] $T_{res}$
$=\frac{\sqrt{s}\Gamma_{res}(s)/\rho_{\pi\pi}}{M^2_{res}-s+
\mbox{Re}\Pi_{res}(M^2_{res})-\Pi_{res}(s)}$
$=\frac{e^{2i\delta_{res}}-1}{2i\rho_{\pi\pi}}$,\ \
 $M^2_{res}$=$m_\sigma^2-\mbox{Re}\Pi_{res}(M^2_{res})$,\ \
 \\[5pt]
$T^2_0$=$\frac{T_0^{2(tree)}}{1-i\rho_{\pi\pi}T_0^{2(tree)}}$=$
\frac{e^{2i\delta_0^2}-1}{2i\rho_{\pi\pi}}$.\\[5pt]
 The
chiral shielding of the $\sigma(600)$ meson in
$\pi\pi$\,$\to$\,$\pi\pi$ is shown  in Fig.\,2 with  the $\pi\pi$
phase shifts $\delta_{res}$, $\delta_{bg}$, $\delta^0_0$ (a) and
 the corresponding cross sections (b). \vspace*{35mm}
%%%%%%%% Fig. 1 %%%%%%%%%%%%%%%%%%%%%%%%%%%%%%%%%%%%%%%%%%%%%%%%%%%%%%%%%%%%%%%%
\begin{wrapfigure}[9]{R}{85mm}
\centering \vspace*{-6mm}
\hspace*{-45pt}\includegraphics[width=65mm]{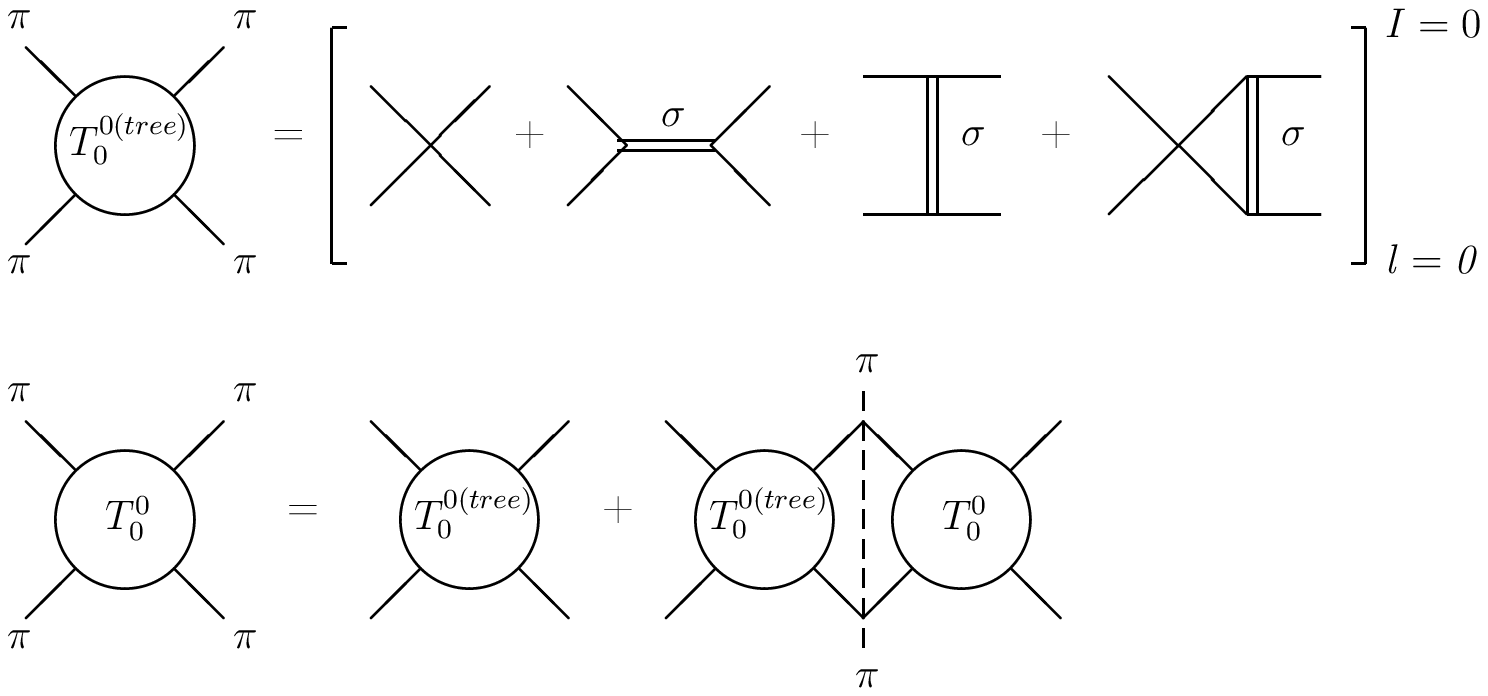}
\caption{ Our approximation.
 } \label{fig1}
\end{wrapfigure}
%%%%%%%%%%%%%%%%%%%%%%%%%%%%%%%%%%%%%%%%%%%%%%%%%%%%%%%%%%%%%%%%%%%%%%%%%%%%%%%%
%%%%%%%% Fig. 2 %%%%%%%%%%%%%%%%%%%%%%%%%%%%%%%%%%%%%%%%%%%%%%%%%%%%%%%%%%%%%%%%
\vspace*{-30mm}
\begin{wrapfigure}[0]{R}{65mm}
\centering \vspace*{-47mm}
\includegraphics[width=65mm]{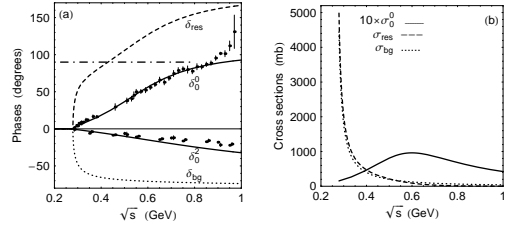}
\caption{$\delta^0_0=\delta_{res}+\delta_{bg}$.} \label{fig2}
\end{wrapfigure}

\vspace{3mm}

\noindent\textbf{3\ \ The \boldmath{$\sigma$} Propagator
\cite{AS9407}}
 \vskip 2mm
$1/D_\sigma(s)$=$1/[M^2_{res}$--$s$+$\mbox{Re}\Pi_{res}(M^2_{res})$--$
\Pi_{res}(s)]$. The $\sigma$ meson self-energy $\Pi_{res}(s)$ is
caused by the intermediate $\pi\pi$ states, that is, by the
four-quark intermediate states. This contribution shifts the
Breit-Wigner (BW) mass greatly
$m_\sigma-M_{res}\approx$\,0.50\,GeV. So, half the BW mass is
determined by the four-quark contribution at least. The imaginary
part dominates the propagator modulus in the region
0.3\,GeV\,$<\sqrt{s}<$\,0.6\,GeV. So, the $\sigma$ field is
described by  its four-quark component at least in this energy
(virtuality) region.

\vspace{2mm}

\noindent\textbf{4\ \ Four-quark Model}

The nontrivial nature of the well-established light scalar
resonances $f_0(980)$ and $a_0(980)$ is no longer denied
practically anybody. As for the nonet as a whole, even a cursory
look at PDG Review gives an idea of the four-quark structure of
the light scalar meson nonet, $\sigma(600)$, $\kappa(700-900)$,
$f_0(980)$, and $a_0(980)$, inverted in comparison with the
classical $P$ wave $q\bar q$ tensor meson nonet $f_2(1270)$,
$a_2(1320)$, $K_2^\ast(1420)$, $\phi_2^\prime(1525)$. Really,
while the scalar nonet cannot be treated as the $P$ wave $q\bar q$
nonet in the naive quark model, it can be easy understood as the
$q^2\bar q^2$ nonet, where $\sigma$ has no strange quarks,
$\kappa$ has the $s$ quark, $f_0$ and $a_0$ have the $s\bar s$
pair. Similar states were found by Jaffe in 1977 in the MIT bag
\cite{Ja77}.

\vspace{2mm}

\noindent\textbf{5\ \ Radiative Decays of the \boldmath{$\phi$}
Meson and the $K^+K^-$ Loop Model \cite{A8907}}

Ten years later we showed that $\phi$\,$\to$\,$\gamma
a_0$\,$\to$\,$\gamma\pi\eta$ and $\phi$\,$\to$\,$\gamma
f_0$\,$\to$\,$ \gamma\pi\pi$ can shed light on the problem of the
$a_0(980)$ and $f_0(980)$ mesons. Now these decays are studied not
only theoretically but also experimentally. When basing the
experimental investigations, we suggested one-loop model $\phi\to
K^+K^-\to\gamma a_0/f_0$, see Fig. 3. This model is used in the
data treatment and is ratified by experiment, see Fig. 4. Gauge
invariance gives the conclusive arguments in favor of the $K^+K^-$
loop transition as the principal mechanism of the $a_0(980)$ and
$f_0(980)$ meson production in the $\phi$ radiative decays.

%%%%%%%% Fig. 3 %%%%%%%%%%%%%%%%%%%%%%%%%%%%%%%%%%%%%%%%%%%%%%%%%%%%%%%%%%%%%%%%
\begin{wrapfigure}[5]{R}{65mm}
\centering \vspace*{-7mm}
\begin{tabular}{ccc}\includegraphics[width=1.8cm]{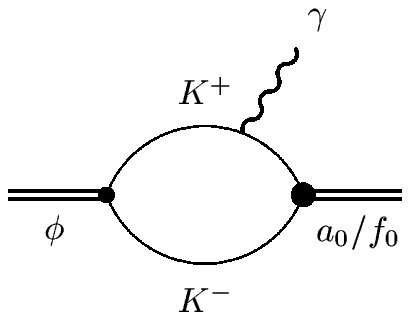}&
\raisebox{-3mm}{$\includegraphics[width=1.8cm]{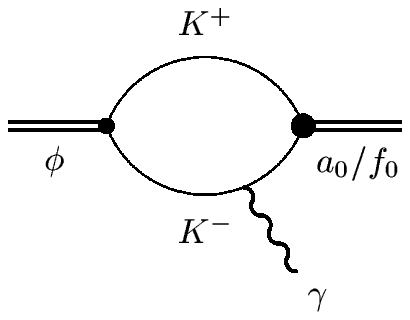}$}&
\includegraphics[width=1.8cm]{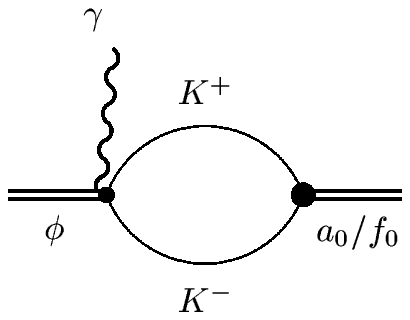}
\end{tabular}\caption{The $K^+K^-$ loop model.}\label{fig3}
\end{wrapfigure}
%%%%%%%%%%%%%%%%%%%%%%%%%%%%%%%%%%%%%%%%%%%%%%%%%%%%%%%%%%%%%%%%%%%%%%%%%%%%%%%%

\vspace{2mm}

\noindent\textbf{6\ \ \boldmath{The $K^+K^-$} Loop Mechanism \\
\hspace*{12pt} is Four-Quark Transition \cite{A8907}}

In truth this means that the $a_0(980)$ and the $f_0(980)$ are
seen in the $\phi$ meson radiative decays owing to the $K^+K^-$
intermediate state. So, the mechanism of the $a_0(980)$ and
$f_0(980)$ production in the $\phi$ meson radiative decays is
established at a physical level of proof. We are dealing with the
four-quark transition. A radiative four-quark transition between
two $q\bar q$ states requires creation and annihilation of an
additional $q\bar q$ pair, i.e., such a transition is forbidden by
the OZI rule, while a radiative four-quark transition between
$q\bar q$ and $q^2\bar q^2$ states requires only creation of an
additional $q\bar q$ pair, i.e., such a transition is allowed by
the OZI rule. The large $N_C$ expansion supports this conclusion.
%%%%%%%% Fig. 4 %%%%%%%%%%%%%%%%%%%%%%%%%%%%%%%%%%%%%%%%%%%%%%%%%%%%%%%%%%%%%%%%
\begin{wrapfigure}[12]{C}{90mm}
\centering \vspace*{-3mm}
\begin{tabular}{cc}\includegraphics[width=7pc,height=5.6pc]{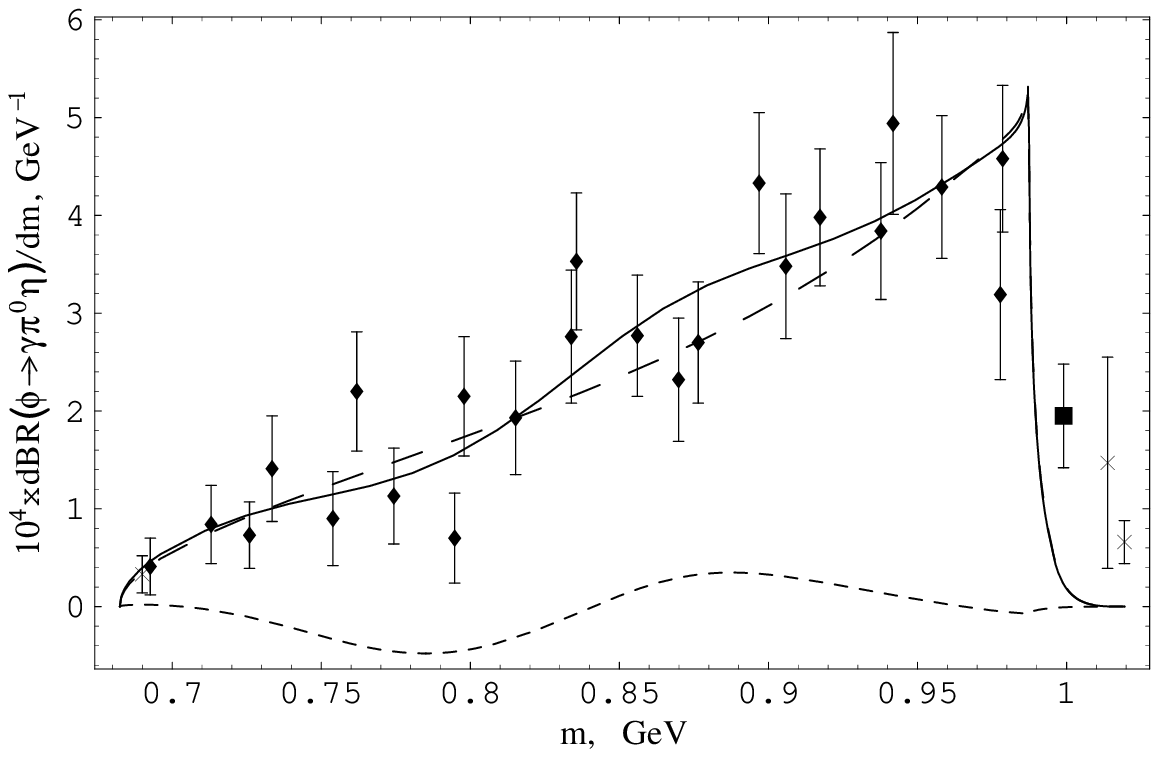}&
{$\includegraphics[width=7pc,height=5.6pc]{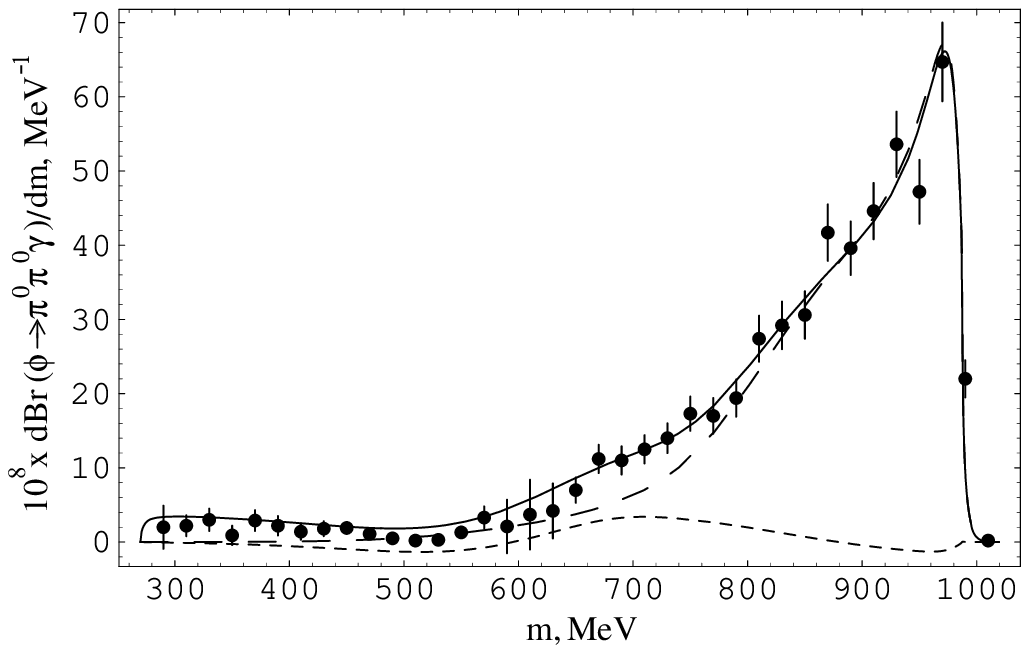}$}
\end{tabular}\caption{The left (right) plot shows the fit
to the KLOE data for the $\pi^0\eta$ ($\pi^0 \pi^0$) mass spectrum
in the $\phi\to\gamma\pi^0\eta$ ($\phi\to\gamma\pi^0\pi^0$) decay
caused by the $a_0(980)$ ($\sigma(600)+f_0(980)$) production
through the $K^+K^-$ loop mechanism.}\label{fig4}
\end{wrapfigure}
%%%%%%%%%%%%%%%%%%%%%%%%%%%%%%%%%%%%%%%%%%%%%%%%%%%%%%%%%%%%%%%%%%%%%%%%%%%%%%%%
%%%%%%%% Fig. 5 %%%%%%%%%%%%%%%%%%%%%%%%%%%%%%%%%%%%%%%%%%%%%%%%%%%%%%%%%%%%%%%%
\begin{wrapfigure}[13]{R}{65mm}
\centering \vspace*{-65mm}
\begin{tabular}{c}\includegraphics[width=12pc,height=16pc]
{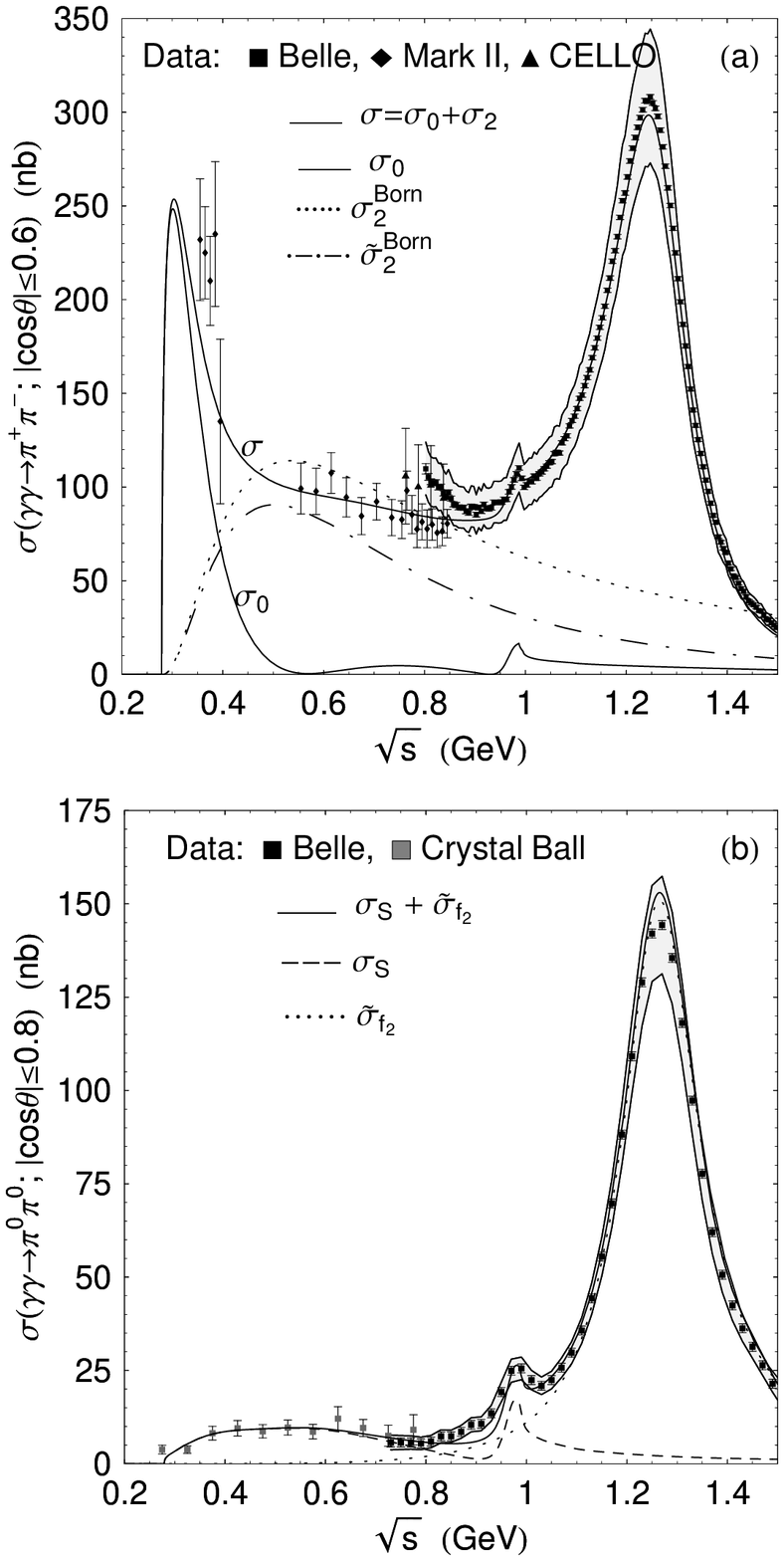}\\
\includegraphics[width=12pc,height=8pc]{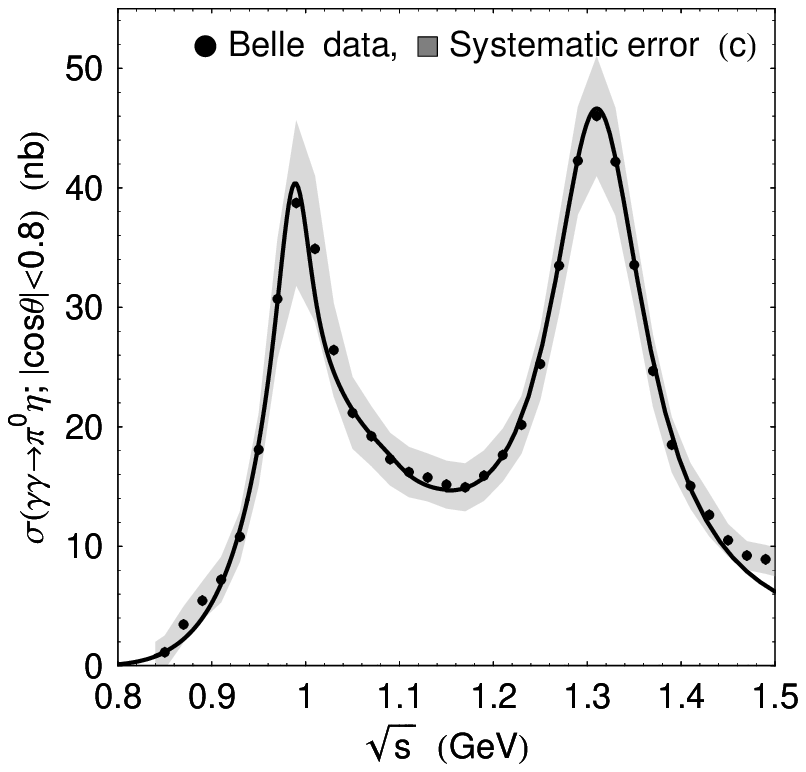}
\end{tabular}\caption{Descriptions of the Belle data on
$\gamma\gamma$\,$\to$\,$\pi^+\pi^-$ (a),
$\gamma\gamma$\,$\to$\,$\pi^0\pi^0$ (b), and
$\gamma\gamma$\,$\to$\,$\pi^0\eta$ (c).}
\label{fig5}\end{wrapfigure}
%%%%%%%%%%%%%%%%%%%%%%%%%%%%%%%%%%%%%%%%%%%%%%%%%%%%%%%%%%%%%%%%%%%%%%%%%%%%%%%%

\vspace{5mm}

\noindent\textbf{7\ \ Scalar Nature and Production\\
\hspace*{11pt}  Mechanisms in \boldmath{$\gamma\gamma$} collisions
\cite{A8209}}

 \vspace{1mm}

  Twenty seven years ago we predicted the
suppression of $a_0(980)\to\gamma\gamma$ and
$f_0(980)\to\gamma\gamma$ in the $q^2\bar q^2$ MIT model,
$\Gamma_{a_0\to\gamma\gamma}\sim\Gamma_{f_0\to\gamma\gamma}\sim
0.27\,\mbox{keV}$. Experiment supported this prediction.

Recently the experimental investigations have made great
qualitative advance. The Belle Collaboration  published data on
$\gamma\gamma\to\pi^+\pi^-$, $\gamma\gamma\to\pi^0\pi^0$, and
$\gamma\gamma\to\pi^0\eta$, whose statistics are huge
\cite{Mo07Ue0809}, see Fig. 5. They not only proved the
theoretical expectations based on the four-quark nature of the
light scalar mesons,  but also have allowed to elucidate the
principal mechanisms of these processes. Specifically, the direct
coupling constants of the $\sigma(600)$, $f_0(980)$, and
$a_0(980)$ resonances with the system are small with the result
that their decays into $\gamma\gamma$ are the four-quark
transitions caused by the rescatterings $\sigma(600)$\,$\to
$\,$\pi^+\pi^-$\,$\to $\,$\gamma\gamma$, $f_0(980)$\,$\to
$\,$K^+K^-$\,$\to $\,$\gamma\gamma$ and $a_0(980)$\,$\to
$\,$K^+K^-$\,$\to $\,$\gamma\gamma$ in contrast to the
$\gamma\gamma$ decays of the classic $P$ wave tensor $q\bar q$
mesons $a_2(1320)$, $f_2(1270)$ and $f'_2(1525)$, which are caused
by the direct two-quark transitions $q\bar q$\,$\to
$\,$\gamma\gamma$ in the main. As a result the practically
model-independent prediction of the $q\bar q$ model
$g^2_{f_2\gamma\gamma}:g^2_{a_2\gamma\gamma}=25:9$ agrees with
experiment rather well. The two-photon light scalar widths
averaged over resonance mass distributions
$\langle\Gamma_{f_0\to\gamma
\gamma}\rangle_{\pi\pi}$\,$\approx$\,0.19 keV,
$\langle\Gamma_{a_0\to\gamma \gamma}\rangle_{\pi\eta}$\,$\approx
$\,0.3 keV and $\langle\Gamma_{\sigma
\to\gamma\gamma}\rangle_{\pi\pi}$\,$\approx$\,0.45 keV. As to the
ideal $q\bar q$ model prediction
$g^2_{f_0\gamma\gamma}:g^2_{a_0\gamma\gamma}=25:9$, it is excluded
by experiment.

\vspace{2mm}

\noindent\textbf{8\ \ Summary  \cite{AS9407,A8907,A8209}}

\noindent\textbf{(i)} The mass spectrum of the light scalars,
$\sigma (600)$, $\kappa (800)$, $f_0(980)$, $a_0(980)$, gives an
idea of their $q^2\bar q^2$ structure.\\ \textbf{(ii)} Both
intensity and mechanism of the $a_0(980)/f_0(980)$ production in
the $\phi(1020)$ radiative decays, the $q^2\bar q^2$ transitions
$\phi\to K^+K^-\to\gamma [a_0(980)$ $/f_0(980)]$, indicate their
$q^2\bar q^2$ nature. \\ \textbf{(iii)} Both intensity and
mechanism of the scalar meson decays into $\gamma \gamma$, the
$q^2\bar q^2$ transitions
$\sigma(600)\to\pi^+\pi^-\to\gamma\gamma$ and [$f_0(980) /a_0(980)
$]\,$\to K^+K^-\to\gamma\gamma$, indicate their $q^2\bar q^2$
nature too.\\
 \textbf{(iv)} In addition, the absence of
$J/\psi$ $\to\gamma f_0(980)$, $a_0(980)\rho$, $f_0(980)\omega$ in
contrast to the intensive $J/\psi$ $\to$ $\gamma f_2(1270)$,
$\gamma f'_2(1525)$, $a_2(1320)\rho$, $f_2(1270)\omega$ decays
intrigues against the $P$ wave $q\bar q$ structure of $a_0(980)$
and $f_0(980)$.

\vspace{4mm}  This work was supported in part by the RFFI Grant
No. 07-02-00093 from the Russian Foundation for Basic Research and
by the Presidential Grant No. NSh-1027.2008.2 for Leading
Scientific Schools.

 \vspace{8mm}

\noindent\textbf{Appendix. Source of Spontaneous Breaking of
Symmetry in Quantum Physics }

 \vspace{5mm}

 It is appropriate to mention here that Nikolay Nikolaevich Bogolyubov was the pioneer of
spontaneous breaking of symmetry in quantum physics.

 Usually N.N. Bogolyubov is considered as the first-rate
mathematician,  the first-rate  mechanic,  and the first-rate
physicist. But in our case the inverse order is more correct. Or
rather at such a height the distinction between the mathematician
and the physicist is insignificant. It is remembered, how K.F.
Gauss checked the sum of the angles in  a triangle, formed by
three tops in mountains of Harz.   The genuine brilliant is
Bogolyubov's pioneer work on superfluidity ({\bf 1947
\cite{NNB47}}) that triggered research spontaneous breaking of
symmetry in quantum physics. Bellow is its history in citations.
(NNA translations and comments are my ones.)

{\bf L. Landau. The theory of superfluidity of helium II. J. Phys.
USSR, 5, 71, 1941.

 L.D. Landau.  The theory of superfluidity of
helium II. JETP, 11, 592, 1941 (in Russian).}

NNA translation. "Tisza has suggested to consider helium II as the
degenerate  ideal boze-gas,  supposing  that the atoms, being in
the ground state (the state  with energy equal to zero), move
through a liquid without a friction  both on  vessel walls and on
other part of a liquid. But, such an idea  cannot be recognized as
a satisfactory one. Not to mention that liquid helium has nothing
to do with  the ideal gas,  atoms,  being in the ground state,
would  not behave  as "superfluid" ones at all. On the contrary,
nothing could prevent the atoms, being in a normal state,  to
collide the exited  ones, i.e., at movement through a liquid they
would experience a friction and a "superfluidity" could  be not
even mentioned."

{\bf N. Bogolubov. On the theory of superfluidity.
 J. Phys. USSR 11 (1947) 23 [Acad. Sci. USSR. J. Phys. 11, (1947).
 23-32].

N. N.  Bogolubov. On the theory of superfluidity.
 Proc. Acad. Sci. USSR. Phys. Series,  1947, v. 11, No 1, pp. 77-90 (in
 Russian).}

NNA translation. "We will try to overcome this basic difficulty
({\it a fluid friction, L.D. Landau, NNA}) and to show that under
some conditions in a weakly nonideal Bose-Einshtein gas "  the
degenerate condensate " can move without a friction on elementary
excitations with an enough small speed. It is essential to notice
that in our theory these elementary excitations are collective
effect and cannot be identified with individual molecules."

{\bf F. London. 1948. AMERICAN MATHEMATICAL SOCIETY. MathSciNet
Mathematical Reviews. MR0022177 Bogolubov, N. On the theory of
superfluidity. Acad. Sci. USSR. J. Phys. 11, (1947). 23-32.}

"The object of this paper is an attempt to withdraw from the
"couterblast of objections" raised by L. Landau and others [same
J. 5, 71-90 (1941)] against the Bose-Einstein theory of liquid
helium [Tisza, C.R. Acad. Sci. Paris 207, 1035-1037, 1186-1189
(1938); F. London, Physical Rev.(2) 54, 947-954 (1938)] and to
adopt the point of view of the latter. By using the method of
second quantization the author tries to show that the phenomenon
of superfluidity can be explained on the basis of a theory of the
degeneracy of a nonperfect Bose-Einstein gas of certain
"quasi-particles" representing elementary excitations of a
continuum. The author ignores the fact that there is no
Bose-Einstein condensation for quasi-particles which, like the
excitations,  have no constant particle number. The result is
obtained under the condition of certain approximations, neglecting
second and higher order terms of a quantity called $\vartheta$,
which characterizes the non-commutable part of the quantized wave
function. Although it might be justifiable to consider this
quantity as small and the series in question as rapidly convergent
it nevertheless appears insufficient to drive the entire absence
of any interaction from a consideration of a first order
approximation alone." {\it Reviewed by F. London.  Copyright
American Mathematical Society 1948, 2009.}

{\bf L.D. Landau. On the theory of superfluidity. DAN USSR, 61,
253, 1948 (in Russian).

 L. Landau. On the theory of superfluidity. Phys. Rev. 75 (1949) 884.  Letters to the
 Editor.}

"It is useful to note that N.N. Bogolyubov has succeeded recently,
by an ingenious application of second quantization, in determining
the general form of the energy spectrum of the a Bose-Einstein gas
with a weak interaction between the particles. As it should be,
the "elementary excitations" appear automatically, and their
energy $\epsilon$  as a function of the momentum $p$ is
represented by a single curve, which has  a linear initial
part."\\

{\bf TWENTY SIX  YEAS LATER}\\

 {\bf V.B. Berestetskii .\\ ELEMENTARY
PARTICLES. First ITEP Physics School, Issue I, page 9.\\
 ATOMIZDAT, Moscow -1973 ( In Russian).}

NNA translation. "As an example of the real manifestation of the
Goldstone ({\it 1961 \cite{Gold61}, NNA}) effect in
nonrelativistic quantum mechanics of system of identical particles
one can give the result received in 1947 by Bogolyubov. It lies in
the fact that the lowest energy excitation of nonideal bose-gas
have character phonons. Phonons are massless particles, the
complex field is the wave function of the bose-particle in the
method of the secondary quantization, the conserved charge is the
number of particles, the vacuum is the state of the bose-einstein
condensation, the vacuum value of the field $\eta = \sqrt{n}$,
where $n$ is the density of particles."\\

{\bf THIRTY ONE YEARS LATER}\\

{\bf L.D. Landau and E.M. Lifshitz. \\ THEORETICAL PHYSICS. VOLUME
IX.

 E.M. Lifshitz and L.P. Pitaevskii.\\  STATISTICAL PHYSICS.
PART 2. Theory of Condensed State.\\  CHAPTER III,  page 123
(footnote 1).\\ Moscow "NAUKA" 1978. (In Russian)}

NNA translation.  "1) The  method, expounded below,   is due to
N.N. Bogolyubov (1947). The use of this method to the bose gas by
Bogolyubov was the first example of the consistent microscopic
finding of the energy spectrum of "quantum fluids"."

NNA: {\bf As far as could be judged from   such an
acknowledgement, Nikolay Nikolaevich Bogolyubov is the author of
the theory of quantum fluids from the first principles.}

\footnotesize{}
\end{document}